\documentclass[twocolumn,showpacs,aps,prl]{revtex4}
\usepackage{epsfig,amsmath,graphicx}

\begin{document}

\title{Spectral Noise Correlations of an Ultrafast Frequency Comb}

\author{Roman Schmeissner, Jonathan Roslund, Claude Fabre, and Nicolas Treps} 
\affiliation{ Laboratoire Kastler Brossel, Sorbonne Universit\'e - UPMC, ENS, Coll\`ege de France, CNRS; 4 place Jussieu, 75252 Paris, France}

\date{\today}

\begin{abstract}

Cavity-based noise detection schemes are combined with ultrafast pulse shaping as a means to diagnose the spectral correlations of both the amplitude and phase noise of an ultrafast frequency comb. The comb is divided into ten spectral regions, and the distribution of noise as well as the correlations between all pairs of spectral regions are measured against the quantum limit. These correlations are then represented in the form of classical noise matrices, which furnish a complete description of the underlying comb dynamics. Their eigendecomposition reveals a set of theoretically predicted, decoupled noise modes that govern the dynamics of the comb. Finally, the matrices contain the information necessary to deduce macroscopic noise properties of the comb. 

\end{abstract}


\maketitle

Ultrafast frequency combs (FC) have found tremendous utility as precision instruments in domains ranging from frequency metrology \cite{udem2002optical,rosenband2008frequency}, optical clocks \cite{diddams2001optical,yeClockReview2014}, broadband spectroscopy \cite{thorpe2006broadband,diddams2007molecular}, and absolute distance measurement \cite{coddington2009rapid,van2012many}. This sensitivity originates from the fact that a comb carries upwards of $\sim 10^{5}$ copropagating, coherently-locked frequency modes \cite{cundiff2003colloquium}. An understanding of the aggregate noise originating from these teeth is essential for assessing the ultimate sensitivity of a given measurement scheme. In practice, FC noise dynamics are typically described with a succinct number of collective properties \cite{haus1993noise,newbury2005theory,newbury2007low}, such as the pulse energy, carrier envelope offset (CEO), or the temporal jitter of the pulse train.  Along these lines, it has been theorized that a variation in one of these parameters perturbs the FC in a manner that consists of adding a particular noise mode to the coherent field structure \cite{haus1990quantum}. Hence, the entirety of the FC noise dynamics is theoretically governed by a set of unique noise modes in which each mode possesses a particular pulse shape \cite{haus1990quantum,haus1993noise}. The existence of such modes would imply a non-negligible role of spectral noise correlations among the individual FC teeth. Although the distribution of noise across a FC has been investigated \cite{bartels2004stabilization,swann2006fiber}, the role of correlations among various frequencies has gone largely unexplored. Toward that end, correlations among disparate spectral bands may be gleaned by observing whether the noise of their spectral sum is equal to the sum of the individual noises. 
Such a scheme may be implemented by combining ultrafast pulse shaping \cite{weiner2000femtosecond}, which furnishes an adjustable spectral filter, with quantum noise-limited balanced homodyne detection \cite{bachor2004guide}. 
This Letter details the extraction of classical amplitude and phase noise matrices for a solid-state femtosecond FC and, in doing so, provides the first experimental realization of uncoupled, broadband noise modes.
Moreover, the identification of spectral correlations offers an enhanced characterization of the comb's noise dynamics from which any collective noise property can be extracted.

The electric field $e(t)$ of the comb is written as a superposition of the constituent teeth: $e(t) = \sum_{n} E_{n}$ in which $E_{n} = A_{n} \exp \left[ i \left( \phi_{n} - \omega_{n} t \right) \right] + \textrm{c.c.} $. The tooth amplitude and phase are represented as $A_{n}$ and $\phi_{n}$, respectively, while the optical frequency $\omega_{n}$ is decomposed as $\omega_{n} = n \, \omega_{\textrm{rep}} + \omega_{\textrm{CEO}}$, in which  $\omega_{\textrm{rep}}$ is the repetition rate of the comb and $\omega_{\textrm{CEO}}$ is the CEO frequency. As FCs are generated within an optical cavity, the coherent structure of this field may be interrupted by various intracavity noise sources, including thermal and mechanical drifts, spontaneous emission, and intensity noise of the pump source. In particular, these perturbations disrupt the field amplitude and phase of a single tooth in a manner described as $ \delta E_{n} = \left( \delta A_{n} + i \delta \phi_{n} A_{n} \right) \cdot \exp \left(  -i \omega_{n} t \right)$ where $\delta E_{n} = E_{n} - \langle E_{n} \rangle$ \footnote{Note that it has tacitly been assumed that all of the comb teeth share a common mean phase of $\langle \phi_{n} \rangle = 0$ (i.e., a transform-limited pulse), which implies that the quadratures containing the phase fluctuations are aligned for all frequencies.}. These fluctuations may be decomposed in terms of their amplitude and phase components to yield 
\begin{equation} \label{eq:homodyne}
\operatorname{Re}(\delta E_{n}) = \delta A_{n} \cos (\omega_{n} t) + A_{n} \delta \phi_{n} \sin(\omega_{n} t).
\end{equation}
Thus, variations of the spectral amplitude are manifest in the same quadrature as that carrying the mean electric field while phase fluctuations are observed in the conjugate quadrature. Homodyne detection offers a quantum noise-limited, phase-sensitive detection scheme capable of measuring these quadrature-dependent fluctuations.
In order to characterize the frequency distribution of noise, the collective parameters of pulse energy and phase described above are conferred a spectral dependence. 

\begin{figure}[htbp]
\centering
\includegraphics[width=70mm]{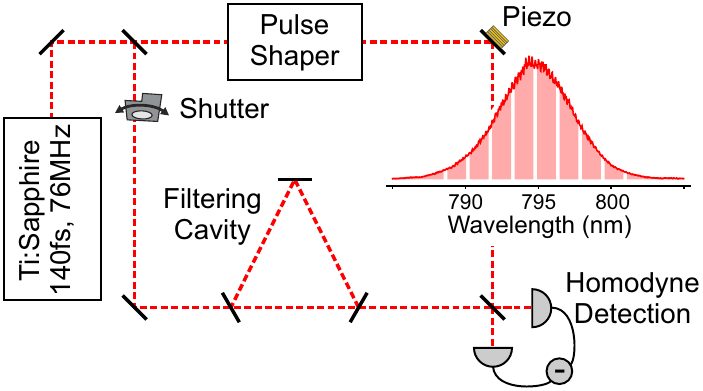}
\caption{Experimental layout for characterizing the classical noise in optical frequency combs. A titanium-sapphire oscillator produces a 76MHz train of $\sim 140 \textrm{fs}$ pulses centered at 795nm. This source is split, and one portion is directed to a 512 element, programmable 2D liquid-crystal pulse shaper, which serves to select a certain spectral region for noise analysis. When the shutter is closed, the balanced homodyne detection measures intensity noise in the spectral slice transmitted by the pulse shaper. Conversely, when the shutter is open, a part of the source is filtered in a Fabry-Perot filtering cavity (finesse of $\mathcal{F} = 420$), which attenuates high frequency fluctuations. This filtered beam is analyzed with homodyne detection, in which the shaped beam serves as the local oscillator. Phase variations are analyzed by locking the relative phase between the two interferometer arms with a piezo-controlled mirror.}
\label{fig-exp-setup}
\end{figure}

A titanium-sapphire oscillator is utilized in which neither the CEO nor the repetition rate is locked. 
This source is divided, and one part constitutes a reference field, which is necessary for homodyne detection, while the other comprises the signal beam to be analyzed. 
This signal field is delivered to a programmable 512-element 2D liquid-crystal pulse shaper, which is capable of independent amplitude and phase modulation \cite{vaughan2005diffraction}. The pulse shaper partitions the $\sim 6 \textrm{nm}$ FWHM spectrum into ten non-overlapping bands of equal width ($\sim 1.5 \textrm{nm}$ per band). This segmentation allows the spectral amplitude and phase noise levels to be interrogated for each spectral band as well as all possible pairs of bands.


The reference beam and the signal constitute two arms of a Mach-Zehnder interferometer as seen in Fig.~\ref{fig-exp-setup}. Since these two fields share a common phase, they must be decoupled in some manner in order to observe the phase noise of the laser source. This decoupling is achieved by introducing the reference beam into a high-finesse ($\mathcal{F} \simeq 420$) Fabry-Perot filtering cavity. The cavity has a free spectral range equal to that of the pulse train repetition rate (i.e., $76 \textrm{MHz}$), and its length is locked with a Pound-Drever-Hall scheme. Sideband fluctuations of the reference field transmitted by the cavity are strongly attenuated for frequencies higher than the cutoff frequency of $\sim 90 \textrm{kHz}$ \cite{Schmeissner2014}, whereas the initial fluctuations persist in the signal arm of the interferometer \footnote{In a traditional homodyne setup, the local oscillator (LO) field is significantly stronger than the reference field, i.e., $A_{\textrm{LO}} \gg A_{\textrm{ref}}$, which allows for measuring the fluctuations of the weaker reference field $E_{\textrm{ref}}$. In the present setup, however, the fluctuations are present on the stronger signal field, which serves as the LO, while they are attenuated on the reference field. Yet, since the interferometer measures the relative phase between its two arms, these fluctuations may be taken to originate in either arm. Thus, it is important to note that the present implementation of homodyne detection is equivalent to the traditional one.}. These two fields are recombined on a 50:50 beamsplitter and then detected with a pair of balanced silicon photodiodes. A mirror mounted on a piezo stack in one arm of the interferometer permits locking the mean relative phase between the two interferometer arms to a value of $\pi / 2$ (i.e., the phase quadrature). As homodyne detection is a projective technique, the difference of the photocurrents for the two diodes provides the quadrature noise of the field projected onto the spectrally-shaped signal. In particular, this difference is directly proportional to the relative phase $\phi_{\textrm{rel}} = \phi_{\textrm{sig}} - \phi_{\textrm{ref}}$ between the two interferometer arms. Since the fluctuations of $\phi_{\textrm{ref}}$ are significantly attenuated, this measure provides the phase noise of the field, i.e., $\delta \phi_{\textrm{rel}} \simeq \delta \phi_{\textrm{sig}}$ . Additionally, the photocurrent difference obtained in the absence of the weaker reference field provides the shot noise level, which normalizes the observed phase fluctuations. This experimental arrangement is conceptually similar to that found in Ref.~\cite{roslund2013wavelength}.

\begin{figure}[htbp]
\centering
\includegraphics[width=85mm]{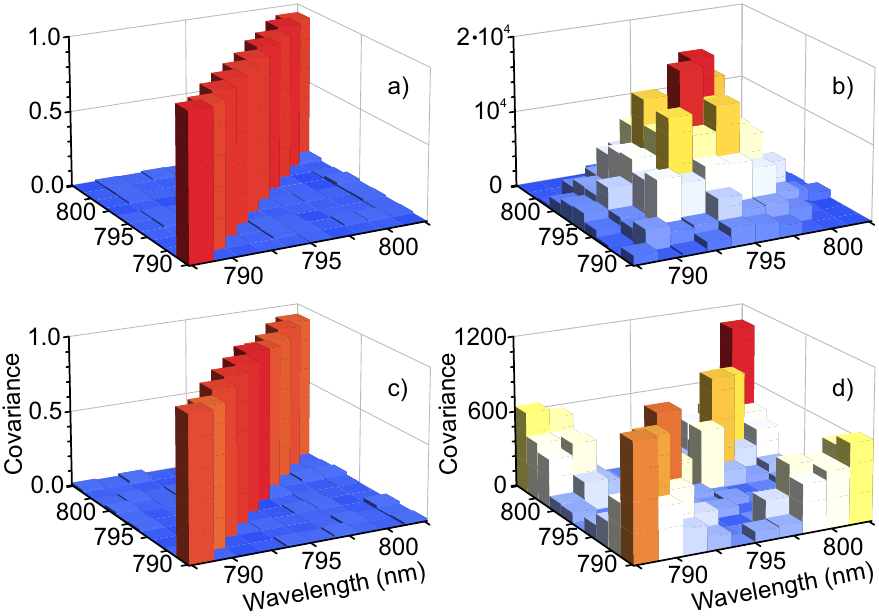}
\caption{Classical covariance matrices for the amplitude and phase quadratures of a titanium-sapphire frequency comb. The matrices are shown relative to shot noise, which has a value of 1.0. Panels (a) and (b) display the phase noise at 3MHz and 500kHz, respectively, while panels (c) and (d) depict the amplitude quadrature at analogous RF analysis frequencies. Both field quadratures exhibit shot noise limited dynamics at 3MHz. Conversely, quadrature-dependent, correlated noise is evident for longer analysis timescales.}
\label{fig-matrices}
\end{figure}

Amplitude fluctuations of the field are subsequently assessed by blocking the reference beam of the interferometer with a shutter and examining the sum of the two photocurrents, which is directly proportional to the signal field's intensity noise. The corresponding shot noise level is retrieved by measuring the difference of the photocurrents for the two diodes. By normalizing the intensity fluctuations to the appropriate shot noise levels, the observed intensity variations are converted to field amplitude noise \cite{opatrny2002mode}. Photocurrent fluctuations arising from both amplitude and phase noise of the source are analyzed with a spectrum analyzer in the sideband RF frequency range of $20 \textrm{kHz} - 5 \textrm{MHz}$.

The observed noise fluctuations closely follow Gaussian statistics, which enables their dynamics to be described with a moment-based covariance matrix. In order to do so, a quadrature operator $\mathcal{O}_{i}$ is associated with each of the ten spectral zones created by the pulse shaper, i.e., $i = 1 \ldots 10$. In line with Eq.~\ref{eq:homodyne}, this operator assumes the form $\mathcal{O}_{i} = \delta \tilde{A}_{i}$ for measurements of the amplitude noise (i.e., when the filtering cavity is blocked) whereas its value for measurements of the phase quadrature is given by $\mathcal{O}_{i} =  \tilde{A}_{i} \, \delta \tilde{\phi}_{i}$, where $\tilde{A}_{i}$ and $\tilde{\phi}_{i}$ are the amplitude and phase of each spectral partition, respectively. The spectrum analyzer measures the RF power of the photocurrent fluctuations, which therefore provides a measure of the noise variance $\langle \mathcal{O}_{i}^{2} \rangle$. Covariance matrices for the amplitude and phase quadratures are independently assembled from 55 distinct measurements for each quadrature, and the noise covariance between two spectral regions is assessed as \cite{spalter1998observation}:
\begin{multline}
\langle  \mathcal{O}_{i} \mathcal{O}_{j} \rangle  = \left[ \langle (\mathcal{O}_{i} + \mathcal{O}_{j})^2 \rangle_{M} - \frac{P_{i}}{P_{t}} \langle \mathcal{O}_{i}^2 \rangle_{M} \right. \\ \left. - \frac{P_{j}}{P_{t}} \langle \mathcal{O}_{j}^2 \rangle_{M} \right] \times \frac{P_{t} }{2 \sqrt{P_{i} \, P_{j}}}    ,
\end{multline}
where $\langle \mathcal{O} ^{2} \rangle_{M}$ is a measured quadrature variance, $P_{i}$ is the optical power associated with a given spectral slice of the comb (evaluated with the mean photodiode signal), and $P_{t} = P_{i} + P_{j}$.

The assembled covariance matrices for both the amplitude and phase quadratures are shown in Fig.~\ref{fig-matrices} for two different interrogation timescales. All of the matrices display the covariance elements relative to the shot noise limit, which possesses a value of 1.0. For high frequency sideband fluctuations ( $\sim 3 \textrm{MHz} $), the noise in each spectral band is identical and equal to the shot noise limit (panels (a) and (c) of Fig.~\ref{fig-matrices}). This is expected since the source itself is shot noise limited for sideband frequencies larger than several megahertz. Moreover, neither the amplitude nor phase matrix exhibit correlations among disparate optical wavelengths for high analysis frequencies, which is in accord with the fact that vacuum fluctuations are entirely uncorrelated for different frequency modes. For longer analysis timescales, however, the matrices are no longer diagonal, and correlations among various frequencies become readily apparent (panels (b) and (d) of Fig.~\ref{fig-matrices}). Amplitude quadrature noise and correlations are predominantly localized in the spectral wings. Conversely, the noise and correlations for the phase quadrature are largely confined to the spectral center. The set of amplitude and phase noise matrices is qualitatively similar across the range of considered RF frequencies, although the magnitude of the covariance values increase with an increasing analysis timescale. 

\begin{figure}[htbp]
\centering
\includegraphics[width=85mm]{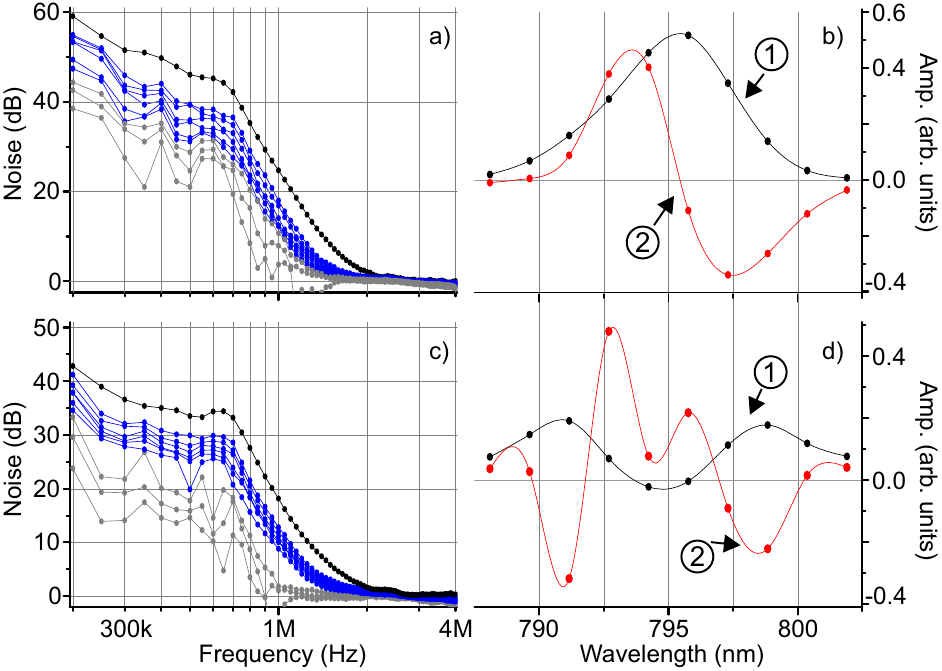}
\caption{Eigendecomposition of the classical amplitude and phase quadrature covariance matrices. The noise eigenvalues for the phase and amplitude quadratures are shown in panels (a) and (c), respectively. A dominant eigenmode is evident for both quadratures. The leading two eigenmodes at a RF frequency of $\sim 1\textrm{MHz}$ are shown in panels (b) and (d) for the phase and amplitude quadrature, respectively. A spline interpolation has been added for visualization purposes.}
\label{fig-eigen}
\end{figure}

One of the primary aims of this work is to discover an underlying modal representation for the FC fluctuations, and such a description becomes accessible upon eigendecomposition of the quadrature covariance matrices. The eigenvalues for both quadratures are dependent upon the analysis frequency and are shown in Fig.~\ref{fig-eigen}. Above a frequency of $\sim 2 \textrm{MHz}$, the noise in both quadratures is equal to that of vacuum fluctuations, which originate from the quantum nature of light. 
Conversely, the individual eigenvalues rise above the shot noise limit for longer analysis timescales but do so in a nondegenerate fashion as seen in Fig.~\ref{fig-eigen}a,c. The noise is also observed to generally be higher in the phase quadrature. Importantly, a principal noise mode is evident in both quadratures and accounts for approximately $\sim 60 \%$ and $\sim 45 \%$ of the fluctuations in the amplitude and phase quadrature, respectively. The existence of a principal eigenmode also accounts for the fact that the covariance matrices of both quadratures exhibit purely positive correlations \cite{opatrny2002mode}.

The eigenmodes corresponding to these eigenvalues reveal the spectral composition of the observed noise correlations. These structures are approximately frequency independent for detection frequencies $\lesssim 2 \textrm{MHz}$. The leading two modes for each quadrature are shown in Fig.~\ref{fig-eigen}b,d for a detection frequency of $\sim 1 \textrm{MHz}$. In the case of the phase quadrature, the leading noise mode is very similar to the spectral envelope of the mean field, 
whereas the secondary eigenmode is comparable to the spectral derivative of this mean field envelope. 
Consequently, to a first approximation, the field is described with a fixed amplitude that is subject to fluctuations of a common phase, which is consistent with a traditional picture of field noise. Additionally, these two eigenstructures closely approximate the theoretical phase quadrature noise modes originally predicted by Haus, \emph{et al.} \cite{haus1990quantum}.
The amplitude quadrature eigenmodes, on the other hand, are markedly different and not amenable to straightforward interpretation. Nonetheless, it is worth mentioning that the leading amplitude mode seen in Fig.~\ref{fig-eigen}d resembles the theoretical mode that characterizes pulse energy fluctuations \cite{haus1990quantum}.



It is also possible to derive the comb's collective noise parameters given knowledge of the noise distribution on the underlying optical teeth. In particular, each collective parameter possesses a specific spectral mode, and its characteristics may be elucidated following an appropriate basis transformation of the comb teeth. 
As a means for demonstrating this capability, the collective properties of CEO phase noise and group delay jitter \cite{paschotta2004noise} are extracted from the covariance matrices shown in Fig.~\ref{fig-matrices}.

The electric field of the comb is rewritten in the form $e(t) = \exp \left[ -i \omega_{0} t \right] \cdot \sum_{n} A_{n} \exp \left[ \phi_{n} - \Omega_{n} t \right]$, where $\omega_{0}$ is the optical carrier frequency and $\Omega_{n} = \omega_{n} - \omega_{0}$. By identifying a slowly-varying complex field envelope as $E(t) = \sum_{n} A_{n} \exp \left[ \phi_{n} - \Omega_{n} t \right]$, the field assumes the simplified form $e(t) = E(t) \cdot \exp \left[ -i \omega_{0} t \right] $.
Fluctuations in the pulse arrival time $\delta \tau$ are then considered, which induce a corresponding variation of the field $ \delta e(t,\delta \tau)$. For temporal variations $\delta \tau$ smaller than the duration of the optical period, i.e., $\delta \tau \ll 1 / \omega_{0}$, the field may be expanded to first order in time, which enables the perturbation to be written $\delta e(t,\delta \tau) = e(t+\delta \tau) - e(t) \simeq \delta \tau \cdot \partial e(t) / \partial t$ \cite{von1986characterization,paschotta2006optical,lamine2008quantum}. Given the decomposition of the field into an envelope and the optical carrier, this perturbation is shown to be $\delta e(t, \delta \tau) = \delta \tau \cdot \left[ -i \omega_{0} E(t) + \partial E(t) / \partial t \right] \cdot \exp \left[ - i \omega_{0} t \right]$. The first term encapsulates field noise that arises from variations of only the optical carrier and corresponds to CEO noise. Conversely, the second term expresses field noise resultant from fluctuations of the envelope arrival time, i.e., repetition rate jitter. 

The temporal field variation $\delta e(t, \delta \tau)$ is then Fourier-transformed in order to reveal the modal representation for these two noise sources, i.e., $\delta E(\omega,\delta \tau) = (1 / 2 \pi) \cdot \int_{-\infty}^{\infty} dt \, \delta e(t, \delta \tau) \cdot \exp[i \omega t]$. Upon doing so, the spectral representation of the field fluctuations is written as
\begin{eqnarray} \label{eq:timing-modes}
\delta E(\omega) &\simeq& -i \, \delta \tau \left[ \omega_{0} \, E(\Omega) + \Omega \cdot E(\Omega) \right] \\ \nonumber
& \simeq & -i \, \delta \tau \left[ w_\textrm{CEO} + w_\textrm{rep} \right],
\end{eqnarray}
where the leading spectral mode $w_\textrm{CEO}$ captures CEO noise and the subsequent mode $w_\textrm{rep}$ describes temporal jitter. Importantly, these two spectral modes are orthogonal and are both represented in the phase quadrature. 


\begin{figure}[htbp]
\centering
\includegraphics[width=75mm]{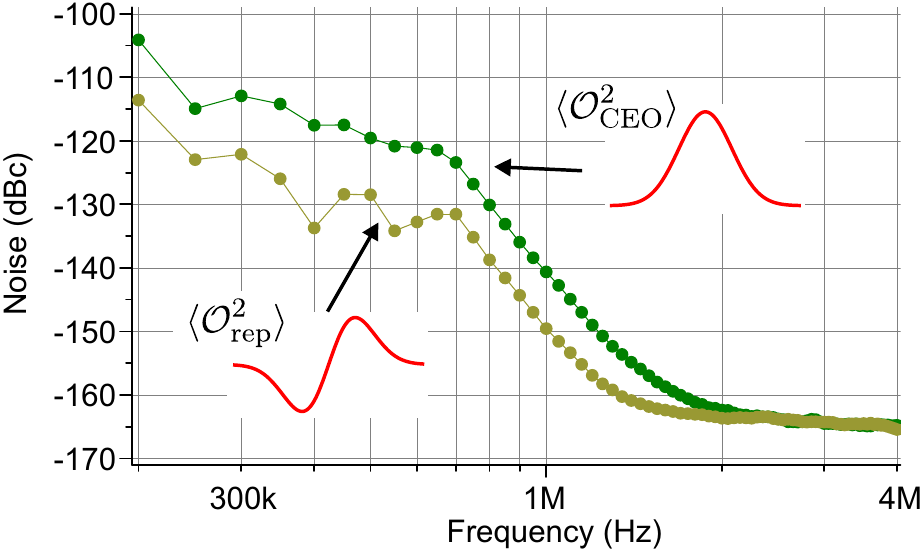}
\caption{Noise variances for CEO phase fluctuations and temporal jitter that are extracted from the classical noise matrices. The CEO noise is the dominant noise property of the presently utilized frequency comb.}
\label{fig-projection}
\end{figure}

The modal forms $w_\alpha$, where $\alpha \in \{ \textrm{CEO},\textrm{rep} \}$, are discretized in a manner corresponding to the ten investigated spectral zones, which provides the vectorial representations $\vec{w}_\alpha$. The phase noise variances of these modes, $\langle \mathcal{O} _\alpha^{2} \rangle$, are retrieved by projecting $\vec{w}_\alpha$ onto the phase quadrature eigenmodes $\vec{\psi}_{\textrm{ph},k}$. This inner product is weighted by the corresponding eigenvalue $\sigma_{\textrm{ph},k}^{2}$. The variances are then written as $\langle \mathcal{O} _{\alpha}^{2} \rangle = \sum_{k} \sigma_{\alpha,k}^{2} \left( \vec{\psi}_{\textrm{ph},k} \cdot \vec{w}_{\alpha} \right)^{2}$ where $\alpha \in \{ \textrm{CEO},\textrm{rep} \}$.

The extracted spectral densities are shown in Fig.~\ref{fig-projection} \footnote{See supplemental material}. 
The CEO phase noise levels exceed the noise levels attributed to temporal jitter by approximately 10\,dB; therefore, CEO noise is the dominant noise property of the presently studied FC. Furthermore, the similarity of $\vec{w}_\textrm{CEO}$ and $\vec{w}_\textrm{rep}$ to the leading two eigenmodes of the phase covariance matrix implies that the structure of this matrix is largely dictated by these two noise sources. Importantly, any collective comb property that is expressible as a superposition of the underlying optical frequencies may potentially be deduced from the intrinsic noise matrices.

It should be noted that intra-quadrature correlations of the form $\langle \delta A \, \delta \phi \rangle$ may alter the composition of the phase and amplitude eigenmodes. Since the measured phase noise significantly exceeds the amplitude fluctuations (Fig.~\ref{fig-eigen}a,c), the effect of quadrature coupling on the phase eigenmodes is expected to be minimal. However, such correlations have the potential to contribute significantly to the amplitude quadrature, which may explain why the form of the amplitude eigenmodes is not as clearly connected to theoretical expectations as the phase eigenmodes. Accordingly, knowledge of these intra-quadrature correlations is essential for assembling a truly decoupled modal representation. The presently utilized apparatus may be adapted to directly measure these intra-quadrature correlations.  

In conclusion, correlations in the noise fluctuations are shown to exist among the underlying teeth of an optical FC by combining ultrafast pulse shaping with balanced homodyne detection. Knowledge of these correlations is utilized to construct classical noise covariance matrices, which are introduced as a tool to characterize the 
amplitude and phase noise present in ultrafast FCs. 
These matrices generalize the theoretical methodology originally outlined by Haus \emph{et al.} and provide the first demonstration of uncoupled noise modes in FC sources. 
Importantly, these techniques are not limited to solid-state comb sources and may be implemented for a variety of systems, including both fiber- and microresonator-based FCs for which the noise dynamics are not as thoroughly understood. The ability to diagonalize broadband, correlated classical noise provides an improved understanding of the noise present in FC and its ultimate effect on measurement sensitivity. 


This work is supported by the European Research Council starting grant Frecquam and the French National Research Agency project Comb. C.F. is a member of the Institut Universitaire de France. J.R. acknowledges support from the European Commission through Marie Curie Actions.

\bibliographystyle{apsrev}


\onecolumngrid
\newpage
\twocolumngrid

\section{Supplementary Material}

This supplement details how the experimental setup presented in the main text of the article is utilized to determine noise levels for a given mode of the frequency comb. These noise levels are determined relative to the optical carrier, which enables comparison to other reported FC noise levels in the literature. In particular, the methodology presented in this supplement is exploited to construct Fig.~4 of the main article. 

\section{Phase Noise Decoupling}

In the absence of a filtering cavity, both arms of the interferometer share identical phase fluctuations since they originate from a common source. As a result, the phase noise between the two arms should be perfectly correlated, i.e., $\delta \phi_{\textrm{sig}} = \delta \phi_{\textrm{ref}} = \delta \phi_{\textrm{0}}$, where $\delta \phi_{\textrm{0}}$ are the phase fluctuations of the laser source. However, phase perturbations in the interferometer itself (i.e., mechanical or thermal fluctuations) will disrupt this correlation and induce a relative phase noise at the homodyne detection. Thus, it is important to distinguish noise originating within the interferometer from the intrinsic noise of the laser field. This discrimination is provided based upon the frequency content of the noise. Phase fluctuations originating from the interferometer are generally low frequency ($ \lesssim 10 \textrm{kHz}$) while the present study is concerned with high-frequency ($ \gtrsim 100 \textrm{kHz}$) field noise that arises from within the oscillator cavity. Upon locking the average relative phase between the two arms of the interferometer, the low-frequency noise component is practically eliminated, which allows consideration of the residual high frequency fluctuations that lie outside the bandwidth of the servo system ($\sim 10 \textrm{kHz}$). Henceforth, the discussion of noise refers exclusively to the intrinsic field noise of the frequency comb source. 

Since the high frequency phase fluctuations are common to both arms of the interferometer, the noise of the relative phase detected by the homodyne apparatus is zero, i.e., $\delta \phi_{\textrm{rel}} = \delta \phi_{\textrm{sig}} - \delta \phi_{\textrm{ref}} = 0$. In order to decorrelate the fluctuations between the two interferometer arms, the reference beam is filtered with a high finesse Fabry-Perot cavity, which attenuates its phase fluctuations. In the absence of any amplitude fluctuations, the phase noise emerging from the cavity can be described as $\delta \phi_{\textrm{ref}} = H(f) \cdot \delta \phi_{\textrm{0}}$, in which $H(f)$ is the frequency-dependent transfer function for the optical sidebands. As such, the relative phase noise measured with homodyne detection is described as

\begin{eqnarray}
\langle \delta \phi_{\textrm{rel}} ^{2} \rangle &=& \langle \left[ \delta \phi_{\textrm{sig}} - \delta \phi_{\textrm{ref}} \right]^{2} \rangle \\ \nonumber
&=& \langle \left[ \delta \phi_{\textrm{0}} - H(f) \cdot \delta \phi_{\textrm{0}} \right]^{2} \rangle \\ \nonumber
&=& \left[ 1 - H(f) \right]^{2} \langle \delta \phi_{\textrm{0}}^{2} \rangle.
\end{eqnarray}
For low sideband frequencies, the cavity does not filter the field ($H \rightarrow 1$), and the two arms of the interferometer remain correlated. As a result, the relative phase noise that is measured approaches zero, i.e., $\langle \delta \phi_{\textrm{rel}} ^{2} \rangle \rightarrow 0$. Conversely, for high sideband frequencies, the cavity perfectly filters the incoming phase fluctuations ($H \rightarrow 0$), such that the detected noise of the relative phase reflects that of the intracavity field, i.e., $\langle \delta \phi_{\textrm{rel}} ^{2} \rangle \rightarrow \langle \delta \phi_{\textrm{0}} ^{2} \rangle$.

Upon knowing the explicit form of the transfer function $H(f)$, the measured relative phase noise $\langle \delta \phi_{\textrm{rel}} ^{2} \rangle $ may be corrected in order to yield the intracavity phase noise $\langle \delta \phi_{\textrm{0}} ^{2} \rangle$. 


\section{Cavity Transfer Function}

The field transmitted by the cavity may be written as $E_{\textrm{ref}} = t(f) \cdot E_{\textrm{0}}$, in which $t(f)$ is the cavity's frequency-dependent transmission factor. The complex laser field $E$ may be decomposed in terms of its amplitude and phase quadratures to yield $E = E^{x} + i \, E^{p}$, in which $E^{x}$ and $E^{p}$ represent the amplitude and phase quadrature of the field, respectively. The field transmitted by the cavity may be decomposed in a similar manner:

\begin{eqnarray}
E_{\textrm{ref}}^{x} &=& \operatorname{Re}(t) E_{\textrm{0}}^{x} - \operatorname{Im}(t) E_{\textrm{0}}^{p} \label{convsys} \\ \nonumber
E_{\textrm{ref}}^{p} &=& \operatorname{Im}(t) E_{\textrm{0}}^{x} + \operatorname{Re}(t) E_{\textrm{0}}^{p}.
\end{eqnarray}
Hence, the cavity interconverts the amplitude and phase quadratures of the input field $E_{\textrm{0}}$. The quadrature fluctuations may also be written in terms of the field amplitude $A$ and phase $\phi$ in a manner analogous to that adopted in the main text of the article: $\delta E_{\textrm{0}}^{x} = \delta A$ and $\delta E_{\textrm{0}}^{p} = A \, \delta \phi $.

As seen in Fig.~3a,c of the main text, the measured phase noise significantly exceeds that of the amplitude noise. Additionally, the forms of the amplitude and phase noise matrices seen in Fig.~2b,d (and their respective eigenvectors) are qualitatively different. Consequently, quadrature-interconverted amplitude fluctuations do not contribute to the observed phase noise in a meaningful way and are henceforth neglected. As a result, the phase quadrature of the field transmitted by the cavity is taken to be $E_{\textrm{ref}}^{p} \simeq \operatorname{Re}(t) E_{\textrm{0}}^{p}$, which implies that the real component of the transmission function dictates the level of phase decoupling. 

The transmission function for a Fabry-Perot filtering cavity is given as
\begin{equation} \label{eq:trans1}
t(f) = \frac{t_{1}t_{2} \exp[i \pi \cdot f / f_{\textrm{rep}}]}{1-r_{1}r_{2} \exp[i 2 \pi \cdot f / f_{\textrm{rep}}]},
\end{equation}
where $t_{1}\,(r_{1})$ and $t_{2}\,(r_{2})$ are the field transmission (reflection) coefficients for the input and output coupler, respectively, and $f_{\textrm{rep}}$ the repetition rate of the input frequency comb (equivalent to the free spectral range when the cavity is at resonance). 

For sideband frequencies $f$ that are significantly smaller than the repetition of the pulse train, i.e., $ f / f_{\textrm{rep}} \ll 1$, the transmission function may be simplified to the form:
\begin{equation} \label{eq:trans2}
t(f) \simeq \sqrt{T_{\textrm{max}}} \cdot \frac{1+ i \, \pi \left( \frac{1+r_{1}r_{2}}{1-r_{1}r_{2}} \right) \cdot f / f_{\textrm{rep}} }{1+\pi^{2} \, F \cdot (f / f_{\textrm{rep}})^2 }
\end{equation}
where $T_{\textrm{max}} = \left[ t_{1}t_{2} / (1-r_{1}r_{2}) \right]^{2}$ is the maximum transmission of the field intensity and the coefficient $F = 4 r_{1} r_{2} / (1 - r_{1} r_{2})^{2}$ is related to the cavity finesse $\mathcal{F}$ through the equation $F = 1 / \sin^{2} \left[ \pi / (2 \mathcal{F}) \right]$. For a high finesse cavity (i.e., $\mathcal{F} \gg 1$), Eq.~\ref{eq:trans2} may again be simplified to yield
\begin{equation} \label{eq:trans3}
t(f) \simeq \sqrt{T_{\textrm{max}}} \cdot \frac{1+ i \, (f / f_{c}) }{1+(f / f_{c})^2 },
\end{equation}
where the cutoff frequency $f_{c}$ of the cavity is given as $f_{c} = f_{\textrm{rep}} / (2 \mathcal{F})$.

The transmission factor that attenuates the fluctuations of the field's phase quadrature is therefore finally specified as
\begin{equation}
\operatorname{Re}(t) = \frac{\sqrt{T_{\textrm{max}}}}{1+(f / f_{c})^2}.
\end{equation}
Accordingly, the amplitude of the field emerging from the cavity is given as $A_{\textrm{ref}} = \sqrt{T_{\textrm{max}}} \cdot A_{\textrm{0}}$ while the phase fluctuations are attenuated as $\delta \phi_{\textrm{ref}} = H(f) \cdot \delta \phi_{\textrm{0}}$, where the transfer function is written as:
\begin{equation} \label{eq:transferH}
H(f) = \frac{1}{1+(f / f_{c})^2},
\end{equation}
which is a Lorentzian function specified by its cutoff frequency $f_{c}$.

\subsection{Convergence of Relative Phase Noise}

It is interesting to consider the sideband frequency at which the relative phase noise of the measured signal $\langle \delta \phi_{\textrm{rel}}^{2} \rangle$ is within 3dB of the actual phase noise of the cavity field $\langle \delta \phi_{\textrm{0}}^{2} \rangle$. This frequency is readily found from the relation
\begin{equation}
\left[ 1-H(f_{3 \textrm{dB}}) \right]^{2} = \frac{1}{2}.
\end{equation}
With the form of $H(f)$ specified by Eq.~\ref{eq:transferH}, the 3dB frequency is revealed to be
\begin{eqnarray}
f_{3 \textrm{dB}} &=& f_{c} \cdot \sqrt{\frac{1}{\sqrt{2}-1}} \\ \nonumber
f_{3 \textrm{dB}} &\simeq& 1.55 \cdot f_{c}.
\end{eqnarray}
Thus, for frequencies $f \gtrsim 1.55 f_{c}$, the relative phase noise $\langle \delta \phi_{\textrm{rel}}^{2} \rangle$ that is measured with homodyne detection closely represents the intrinsic phase noise of the laser system $\langle \delta \phi_{\textrm{ref}}^{2} \rangle$.

In the present circumstance fc is 90khz, so above 135khz the noise is predominantly that of the original field. 

\section{Measurement calibration}

\begin{figure}[htbp]
\centering
\includegraphics[width=75mm]{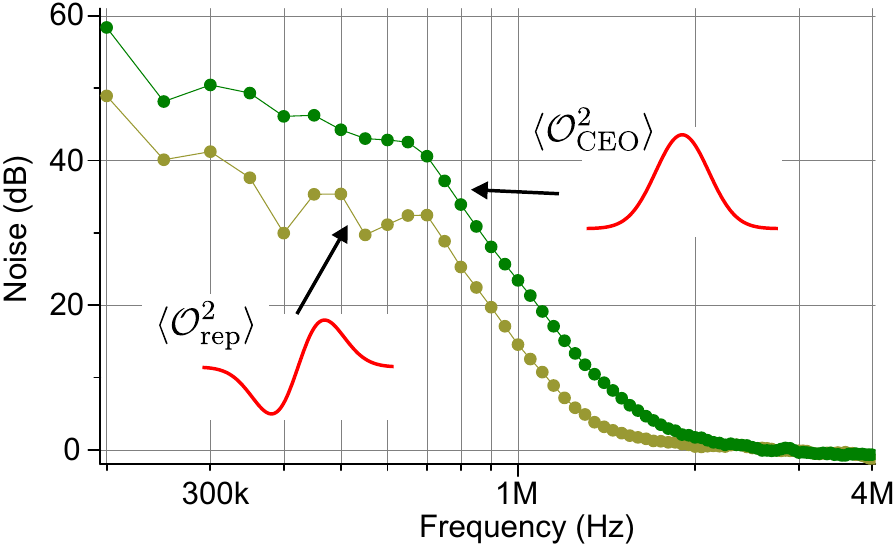}
\caption{Noise variances for CEO phase fluctuations and temporal jitter that are extracted from the classical noise matrices shown in Fig.~2 of the main text. These traces do not include any correction for the filtering effect of the cavity. Additionally, they depict the relevant power spectral densities of the two considered modes relative to the shot noise limit.}
\label{fig-projection-supp}
\end{figure}

For a sufficiently small resolution bandwidth (RBW) of the spectrum analyzer, the noise variance at a given frequency $\langle \mathcal{O}(f) ^{2} \rangle_{\alpha}$ is approximately represented as the product between the power spectral density of the noise fluctuations at that frequency $S_{\alpha}(f)$ and the RBW of the spectrum analyzer, i.e., $\langle \mathcal{O}(f) ^{2} \rangle_{\alpha} \simeq S_{\alpha}(f) \cdot \textrm{RBW}$. The traces shown in Fig.~\ref{fig-projection-supp} depict the CEO phase noise and repetition rate jitter relative to the shot noise limit, i.e., $\langle \mathcal{O}(f) ^{2} \rangle_{\alpha} / \langle \mathcal{O} ^{2} \rangle_{\textrm{shot}} = S_{\alpha}(f) / S_{\textrm{SQL}}$ where $\alpha \in \{ \textrm{CEO},\textrm{rep} \} $.

The standard quantum limit (SQL) is utilized to assess the relative noise magnitude contained in both modes $w_{\alpha}$ to the power of the optical carrier. The SQL is the minimum level of amplitude or phase noise permitted by the quantum nature of light, and it manifests itself as a white noise floor from which the measured phase fluctuations emerge. A beam of average power $P$ possesses a single sideband power spectral density of $S_{\textrm{SQL}} = 2 \, h \nu_{0} / P$ where $h$ is Planck's constant and $\nu_{0}$ is the optical carrier frequency \cite{bachor2004guide}. The normalized noise levels $\langle \mathcal{O}(f) ^{2} \rangle_{\alpha} / \langle \mathcal{O} ^{2} \rangle_{\textrm{shot}}$ are multiplied by $S_{\textrm{SQL}}$ in order to yield the relative intensity noise in the detected beam (which corresponds to the units of dBc/Hz).

Upon correcting the measured spectral densities by the cavity transfer function of Eq.~\ref{eq:transferH} and the shot noise limit $S_{\textrm{SQL}}$, the inferred relative noise is represented in a logarithmic scale as
\begin{equation}
S_{\alpha,\textrm{dB}} = S_{\alpha,\textrm{dB}}^{\textrm{meas}} - F(f)_{\textrm{dB}} + S_{\textrm{SQL,dB}}
\end{equation}
where $S_{\alpha,\textrm{dB}}^{\textrm{meas}}$ are the measured traces of Fig.~\ref{fig-projection-supp}, the filter function $F(f)$ is given as $F(f) = \left[ 1 - H(f) \right]^{2}$, and the subscripts dB indicate that the relevant spectral densities are specified in decibels. The corrected power spectral densities $S_{\alpha,\textrm{dB}}$ for both the CEO and envelope jitter modes are depicted in Fig.~4 of the main text.

\bibliographystyle{apsrev}

\end{document}